\documentclass[conference]{IEEEtran}
\IEEEoverridecommandlockouts
\usepackage{bbding}
\usepackage{xcolor}
\usepackage{colortbl}
\usepackage{multirow}
\usepackage{booktabs}
\usepackage{cite}
\usepackage{amsmath,amssymb,amsfonts}
\usepackage{graphicx}
\usepackage{subcaption}
\usepackage{adjustbox}
\usepackage{textcomp}
\usepackage{xcolor}
\usepackage{xspace}
\usepackage[dvipsnames]{xcolor}
\usepackage{hyperref}
\usepackage{tikz}
\usepackage{svg}
\usepackage[ruled,linesnumbered]{algorithm2e}
\definecolor{cvprblue}{rgb}{0.21,0.49,0.74}
\hypersetup{
    colorlinks=true,
    linkcolor=blue, 
    citecolor=blue, 
    urlcolor=purple 
}
\usepackage{authblk}
\usepackage[capitalize]{cleveref}
\crefname{section}{Sec.}{Secs.}
\Crefname{section}{Section}{Sections}
\Crefname{table}{Table}{Tables}
\crefname{table}{Table}{Tables}
\Crefname{figure}{Figure}{Figures}
\crefname{figure}{Fig.}{Figs.}
\Crefname{equation}{Equation}{Equations}
\crefname{equation}{Eq.}{Eqs.}
\crefname{algocf}{alg.}{algs.}
\Crefname{algocf}{Algorithm}{Algorithms}

\makeatletter
\DeclareRobustCommand\onedot{\futurelet\@let@token\@onedot}
\def\@onedot{\ifx\@let@token.\else.\null\fi\xspace}

\def\etal{\emph{et al}\onedot}
\makeatother
\def\BibTeX{{\rm B\kern-.05em{\sc i\kern-.025em b}\kern-.08em
    T\kern-.1667em\lower.7ex\hbox{E}\kern-.125emX}}
\begin{document}

\title{FS-RWKV: Leveraging Frequency Spatial-Aware RWKV for 3T-to-7T MRI Translation}

\author[1]{Yingtie Lei}
\author[2]{Zimeng Li\thanks{* Corresponding authors: cmpun@umac.mo, xuhangc@hzu.edu.cn}}
\author[1*]{Chi-Man Pun}
\author[3,4]{Yupeng Liu}
\author[1,5*]{Xuhang Chen\thanks{
This work was supported in part by the Science and Technology Development Fund, Macau SAR, under Grant 0193/2023/RIA3 and 0079/2025/AFJ, and the University of Macau under Grant MYRG-GRG2024-00065-FST-UMDF, in part by Shenzhen Medical Research Fund (Grant No. A2503006), in part by the National Natural Science Foundation of China (Grant No. 62501412 and 82300277), in part by Shenzhen Polytechnic University Research Fund (Grant No. 6025310023K), in part by Medical Scientific Research Foundation of Guangdong Province (Grant No. B2025610 and B2023012), and in part by Guangdong Basic and Applied Basic Research Foundation (Grant No. 2024A1515140010).}}

\affil[1]{Faculty of Science and Technology, University of Macau}
\affil[2]{School of Electronic and Communication Engineering, Shenzhen Polytechnic University}
\affil[3]{Department of Cardiology, Guangdong Provincial People's Hospital (Guangdong Academy of Medical Sciences),\protect\\Southern Medical University, Guangzhou, China}
\affil[4]{Guangdong Cardiovascular Institute, Guangdong Provincial People's Hospital,\protect\\Guangdong Academy of Medical Sciences, Guangzhou, China}
\affil[5]{School of Computer Science and Engineering, Huizhou University}

\maketitle

\begin{abstract}
Ultra-high-field 7T MRI offers enhanced spatial resolution and tissue contrast that enables the detection of subtle pathological changes in neurological disorders. However, the limited availability of 7T scanners restricts widespread clinical adoption due to substantial infrastructure costs and technical demands. Computational approaches for synthesizing 7T-quality images from accessible 3T acquisitions present a viable solution to this accessibility challenge. Existing CNN approaches suffer from limited spatial coverage, while Transformer models demand excessive computational overhead. RWKV architectures offer an efficient alternative for global feature modeling in medical image synthesis, combining linear computational complexity with strong long-range dependency capture. Building on this foundation, we propose Frequency Spatial-RWKV (FS-RWKV), an RWKV-based framework for 3T-to-7T MRI translation. To better address the challenges of anatomical detail preservation and global tissue contrast recovery, FS-RWKV incorporates two key modules: (1) Frequency-Spatial Omnidirectional Shift (FSO-Shift), which performs discrete wavelet decomposition followed by omnidirectional spatial shifting on the low-frequency branch to enhance global contextual representation while preserving high-frequency anatomical details; and (2) Structural Fidelity Enhancement Block (SFEB), a module that adaptively reinforces anatomical structure through frequency-aware feature fusion. Comprehensive experiments on UNC and BNU datasets demonstrate that FS-RWKV consistently outperforms existing CNN-, Transformer-, GAN-, and RWKV-based baselines across both T1w and T2w modalities, achieving superior anatomical fidelity and perceptual quality.
\end{abstract}


\section{Introduction}
\begin{figure}[t]
    \centering
    \includegraphics[width=\linewidth]{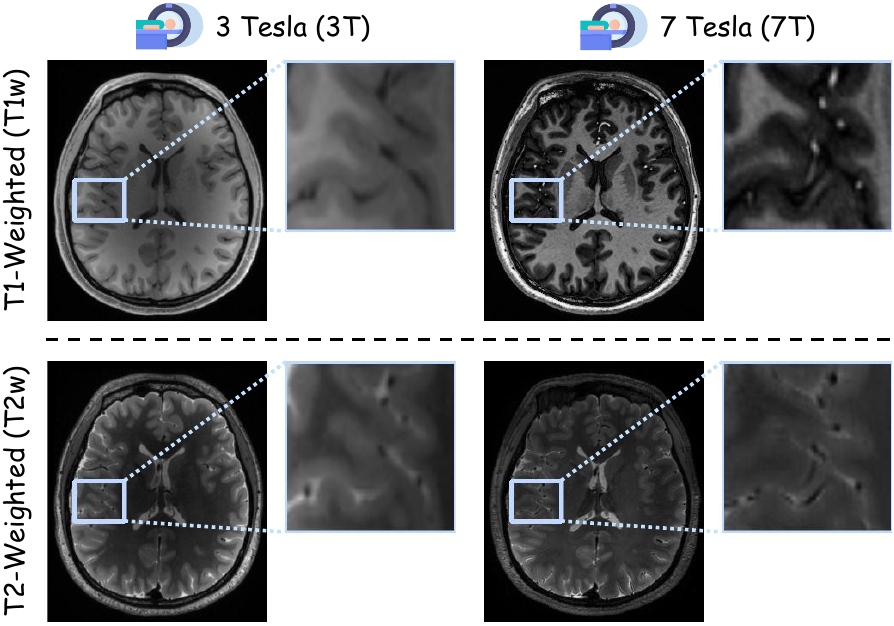}
    \caption{Comparison of T1-weighted and T2-weighted MRI images acquired at 3T and 7T. Images at 7T provide higher spatial resolution, improved tissue contrast, and clearer visualization of anatomical structures compared to 3T.}
    \label{fig:teaser}
\end{figure}

Magnetic resonance imaging (MRI) has become an essential diagnostic tool due to its excellent spatial resolution and non-invasive nature~\cite{westbrook2018mri}. By leveraging strong magnetic fields and radiofrequency waves, MRI enables the visualization of detailed anatomical structures~\cite{van1999basic}. Compared to imaging modalities that rely on ionizing radiation, such as computed tomography (CT) and X-ray imaging, MRI provides superior soft tissue contrast and is safer for repeated clinical use~\cite{aisen1986mri}. These advantages have led to the widespread adoption of MRI in the diagnosis of neurological, musculoskeletal, and cardiovascular diseases~\cite{jack2008alzheimer,gold2004musculoskeletal}. While 3 Tesla (3T) MRI has become a clinical standard, ultra-high-field 7T scanners offer further improvements. Specifically, 7T MRI provides higher signal-to-noise and contrast-to-noise ratios than 3T~\cite{trattnig2018key}, yielding improved spatial resolution and tissue contrast, as shown in~\cref{fig:teaser}. This enhanced sensitivity is particularly valuable in neurological applications, revealing subtle abnormalities in diseases like Alzheimer's and Parkinson's~\cite{van2014cortical,cosottini2014mr}. However, 7T MRI faces practical limitations. The most significant barrier is accessibility, since 7T systems comprise less than 1~$\%$ of the global MRI infrastructure, whereas 3T scanners are much more widely available~\cite{chen2023paired}. Additionally, 7T MRI is more sensitive to motion and susceptible to artifacts caused by field inhomogeneities and susceptibility effects~\cite{barisano2019clinical,vargas2018clinical}. Patients are also more likely to experience discomfort, such as vertigo and dizziness, during 7T examinations~\cite{schaap2016exposure}.

To address these limitations, researchers have developed computational approaches to synthesize 7T MRI from readily accessible 3T data. This strategy aims to achieve ultra-high-field image quality without requiring new hardware or costly infrastructure upgrades. Early work focused on traditional machine learning techniques, including sparse learning~\cite{bahrami2015hierarchical}, random forests~\cite{bahrami20177t}, and linear regression~\cite{zhang2018dual}. While these studies demonstrated the feasibility of computationally approximating superior field strength images, their reliance on handcrafted features limited generalization across diverse clinical scenarios, motivating the adoption of deep learning frameworks.
Convolutional neural networks (CNNs) quickly became popular for this cross-field synthesis task. Bahrami~\etal~\cite{bahrami2016convolutional} demonstrated that CNNs can generate 7T-like contrast and structural detail from 3T scans. Zhang~\etal~\cite{zhang2019dual} proposed a dual-domain network combining spatial and frequency representations, while Qu~\etal~\cite{qu2020synthesized} incorporated wavelet features for enhanced anatomical resolution. Cui~\etal~\cite{cui20247t} validated CNN-based synthesis on traumatic brain injury patients, demonstrating clinical utility.
More recently, vision Transformer architectures have been explored for MRI translation. Eidex~\etal~\cite{eidex2024high} combined CNNs and Transformers to capture local and global features, improving image fidelity. Liu~\etal proposed WFANet‑DDCL~\cite{liu2025wfanet}, integrating wavelet attention with dual-domain learning, achieving competitive performance with limited training data.
\begin{figure}[t]
    \centering
    \includegraphics[width=1\linewidth]{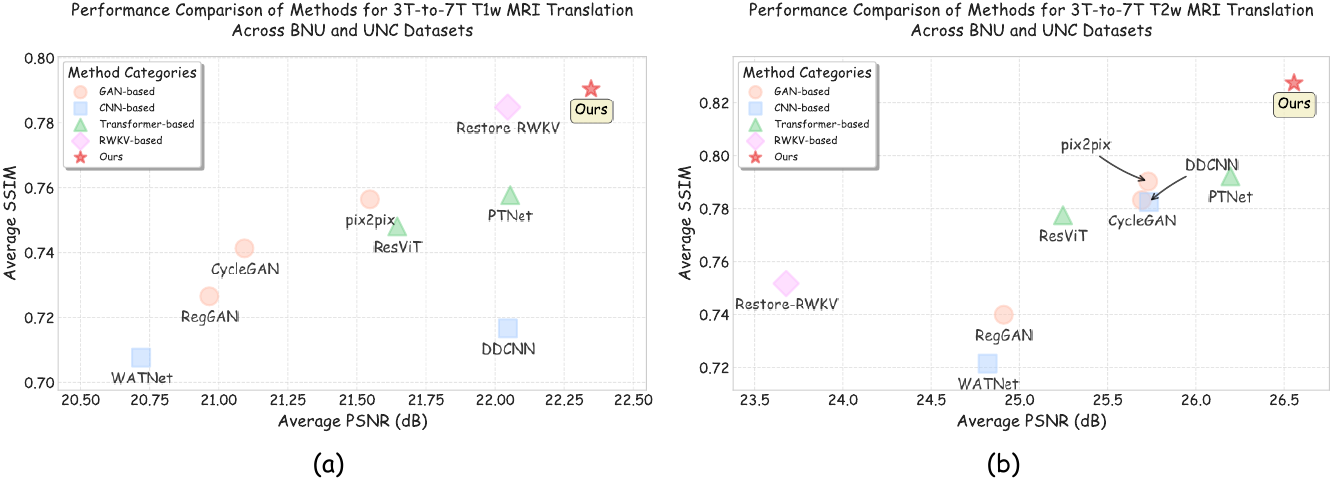}
    \caption{Quantitative comparison of representative methods for 3T-to-7T MRI translation on T1w (a) and T2w (b) images. Each point corresponds to a different method. Our proposed model outperforms existing methods on both modalities.}
    \label{fig:metric-teaser}
\end{figure}

Although CNNs are capable of 7T MRI synthesis to some extent, their limited receptive field restricts the modeling of global anatomical context. Vision Transformers mitigate this with global self-attention, but require substantially more memory and computation~\cite{shamshad2023transformers}. Recently, RWKV has emerged as a promising alternative, offering linear complexity and Transformer-like performance~\cite{peng-etal-2023-rwkv}. Building on this, Vision-RWKV efficiently processes images as token sequences, enabling effective long-range modeling with significantly reduced computational and memory costs compared to traditional Vision Transformers~\cite{duan2025visionrwkv}. Motivated by these observations, we propose Frequency Spatial-Aware RWKV (FS-RWKV), an RWKV-based model for 3T-to-7T MRI translation. Building on insights from~\cite{qu2020synthesized,liu2025wfanet}, our approach integrates frequency-domain information throughout the architecture. Specifically, we introduce a Frequency Spatial Omnidirectional-Shift (FSO-Shift) module that utilizes discrete wavelet transform to decompose feature maps into low-frequency and spatial components, and then applies omnidirectional shifts to each. This design enables the model to jointly capture global tissue contrast and fine anatomical details, both critical for high-fidelity 7T synthesis. In addition, we develop a Structural Fidelity Enhancement Block (SFEB), which adaptively fuses low-frequency, high-frequency, and spatial-domain features. By promoting complementary interactions among these representations, SFEB enhances both tissue contrast and structural detail across multiple resolutions. Together, these architectural innovations contribute to improved synthesis of 7T-quality images, enabling more faithful reconstruction of both broad tissue properties and intricate anatomical details. As illustrated in~\cref{fig:metric-teaser}, our proposed FS-RWKV achieves the highest PSNR and SSIM values among all compared approaches for both T1w and T2w translation tasks. This demonstrates the effectiveness of our architectural innovations in faithfully synthesizing high-quality 7T images.

To summarize, our main contributions are as follows:
\begin{enumerate}
    \item We propose FS-RWKV, a novel architecture that achieves state-of-the-art performance on 3T-to-7T MRI translation, consistently outperforming existing methods on different modalities.
    \item We introduce FSO-Shift, a token shift mechanism that leverages discrete wavelet transform and applies omnidirectional shifts to spatial and low-frequency components, facilitating effective modeling of global context.
    \item We design SFEB to adaptively fuse low-frequency, high-frequency, and spatial-domain features across multiple resolutions, promoting complementary feature interactions and improving the preservation of both tissue contrast and anatomical detail.
\end{enumerate}
\section{Related Work}
3T-to-7T MRI translation aims to synthesize 7T-like images from 3T scans by learning mappings from paired datasets that capture 7T characteristics. Unlike conventional super-resolution focused on upsampling, this involves reconstructing the unique signal and contrast of ultra-high-field MRI, requiring precise recovery of global contrast and detailed anatomical structures. This section reviews approaches for this translation.

\subsection{Traditional Learning-Based Methods}
Before deep learning, researchers explored traditional approaches to reconstruct 7T-like images from 3T MRI using patch-level modeling with sparse representation and dictionary learning. Rueda~\etal~\cite{rueda2013single} developed a super-resolution method using overcomplete dictionaries for high-resolution patches via sparse coding. Bahrami~\etal~\cite{bahrami2015hierarchical} applied a hierarchical strategy with multi-level canonical correlation analysis (CCA) to align 3T and 7T patches in a shared space, followed by group-sparse coding for spatial structure preservation. This was extended~\cite{bahrami2016reconstruction} with adaptive patch sizing and multi-scale fusion. Random forest regression was also explored by Bahrami~\etal~\cite{bahrami20177t} in a two-stage pipeline, while Zhang~\etal~\cite{zhang2018dual} proposed a dual-domain cascaded framework applying patch-wise linear mapping in both spatial and frequency domains.

These traditional methods, while effective in specific settings, remain limited by hand-crafted features and patch-based processing, hindering their ability to model global anatomical structure and complex nonlinear mappings. These constraints motivated the shift toward deep learning approaches.
\subsection{Deep Learning-Based Methods}
Deep learning has introduced more flexible approaches for synthesizing 7T-like MRIs from 3T scans. Early CNN-based efforts by Nie~\etal~\cite{nie2018medical} established the framework for medical image synthesis. Bahrami~\etal~\cite{bahrami2016convolutional} introduced a patch-wise regression model integrating appearance features and anatomical priors, later extending to cascaded CNNs for joint reconstruction and segmentation~\cite{bahrami2017joint}. To improve structural fidelity, Zhang~\etal~\cite{zhang2019dual} proposed a dual-domain network jointly processing spatial and Fourier features, while Qu~\etal~\cite{qu2020synthesized} incorporated wavelet-based decomposition for multi-scale detail preservation. Cui~\etal~\cite{cui20247t} demonstrated clinical relevance by applying CNN-based translation to brain-injury datasets.

Vision Transformers emerged as powerful alternatives for modeling anatomical context across broader spatial extents. Dalmaz~\etal~\cite{dalmaz2022resvit} combined residual Transformer blocks with convolutional layers in an adversarial framework, achieving strong synthesis results across diverse medical imaging tasks. PTNet3D~\cite{zhang2022ptnet3d} extended this with a 3D pyramid Transformer design for longitudinal infant brain MRI synthesis. For 7T MRI synthesis, Eidex~\etal~\cite{eidex2024high} developed a hybrid CNN–Transformer model integrating local features and global anatomical structure to synthesize 7T ADC maps, while WFANet-DDCL~\cite{liu2025wfanet} used wavelet-based frequency attention and dual-domain consistency in a semi-supervised Transformer framework.

Generative models represent another paradigm showing strong potential in medical image synthesis. CycleGAN~\cite{zhu2017unpaired} and pix2pix~\cite{isola2017image} established GANs for image-to-image translation. Dar~\etal~\cite{dar2019image} applied conditional GANs to multi-contrast MRI synthesis, preserving fine details through pixel-wise and adversarial losses. Kong~\etal~\cite{kong2021breaking} addressed challenges in modality consistency and training stability. For 3T-to-7T MRI translation, Duan~\etal~\cite{duan2024synthesized} demonstrated that GAN-generated 7T MPRAGE images closely approximate real clinical scans. Diffusion models have gained attention for high-fidelity synthesis. Özbey~\etal~\cite{ozbey2023unsupervised} introduced SynDiff, combining conditional diffusion processes with cycle-consistency for unsupervised MRI translation. Jiang~\etal~\cite{jiang2023cola} proposed CoLa-Diff, a latent diffusion framework using structural priors to preserve anatomical integrity across modalities.
\section{Proposed Method}
\begin{figure}[ht]
    \centering
    \includegraphics[width=\linewidth]{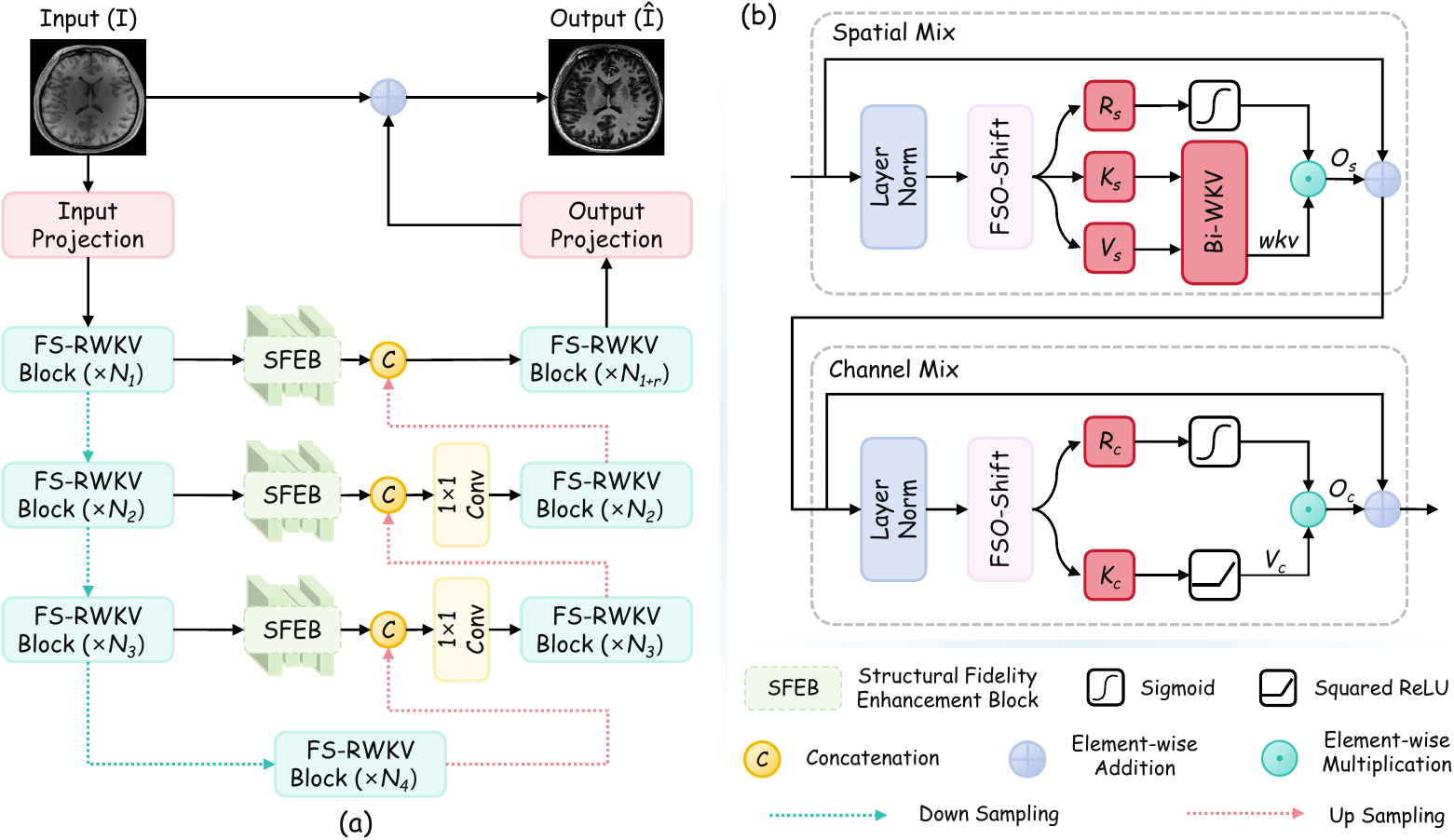}
    \caption{Overview of the FS-RWKV architecture. (a) The overall framework adopts a U-Net structure, where FS-RWKV blocks are used in both encoder and decoder, and SFEB modules enable feature fusion across levels. (b) The detailed design of the FS-RWKV block, which incorporates FSO-Shift modules in both spatial and channel mixing.}
    \label{fig:model}
\end{figure}
\subsection{Overall Architecture}
As illustrated in~\cref{fig:model}, our FS-RWKV adopts a U-Net-style encoder-decoder framework~\cite{ronneberger2015u}, with each stage composed of FS-RWKV blocks and connected via Structural Fidelity Enhancement Blocks (SFEB). The FS-RWKV block leverages FSO-Shift, which applies wavelet decomposition and omnidirectional shifting to model global spatial and frequency interactions. SFEBs are used between encoder and decoder layers to fuse spatial, low-frequency, and high-frequency features across resolutions, enhancing both tissue contrast and anatomical precision. 

\subsection{Frequency Spatial-Aware RWKV (FS-RWKV)}
The RWKV model architecture is defined by four fundamental elements:
\begin{itemize}
    \item R (Receptance): Receives and integrates past information to guide the current representation.
    \item W (Weight): A positional weight decay vector that adaptively modulates the influence of each position, and is learned during training.
    \item K (Key): Encodes features used to identify relevant content, similar to keys in standard attention mechanisms.
    \item V (Value): Represents the content to be propagated or transformed, analogous to values in attention.
\end{itemize}
While the original RWKV block~\cite{peng-etal-2023-rwkv} excels in 1D modeling, its shift mechanism is suboptimal for 2D medical images due to its inability to capture rich spatial context. This is particularly limiting for 3T-to-7T MRI translation, where recovering global tissue contrast and fine anatomical structures is essential. To overcome this, we propose Frequency Spatial Omnidirectional-Shift (FSO-Shift), which performs discrete wavelet decomposition and applies directional shifts to spatial and low-frequency features. This enhances global context modeling, while high-frequency details are reintegrated to preserve anatomical fidelity.

\subsubsection{\textbf{Frequency Spatial Omnidirectional-Shift (FSO-Shift)}}
\begin{figure}[t]
    \centering
    \includegraphics[width=\linewidth]{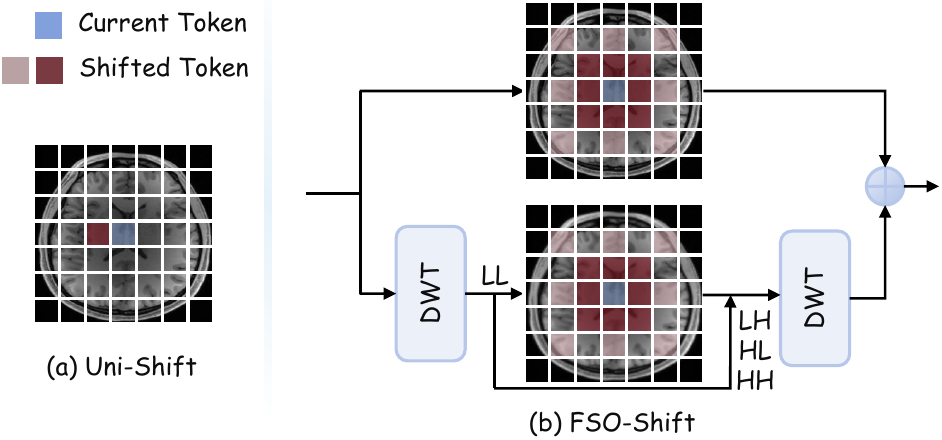}
    \caption{Illustration of (a) Uni-Shift and (b) the proposed FSO-Shift.}
    \label{fig:shift}
\end{figure}

The Frequency Spatial Omnidirectional-Shift (FSO-Shift) module enhances the network's ability to recover global tissue contrast while preserving fine anatomical details in MRI synthesis. As illustrated in~\cref{fig:shift} and detailed in~\cref{alg:fso-shift}, FSO-Shift first applies a discrete wavelet transform to decompose input features into low- and high-frequency components. Inspired by~\cite{chen2025multi}, our design improves flexibility by replacing the limited directional behavior of Uni-Shift~\cite{peng-etal-2023-rwkv} with omnidirectional shifts on both spatial and low-frequency branches, enabling richer contextual interactions.
\begin{algorithm}[t]
\caption{Frequency Spatial Omnidirectional-Shift (FSO-Shift)}
\label{alg:fso-shift}
\KwIn{Input $x$, offset dictionary $D$, learnable weights $\omega_{spatial}, \omega_{LL}, \omega_{out}$}
\KwOut{Shifted output $\mathrm{FSO}\textrm{-}\mathrm{Shift}(x)$}
\tcp{\textbf{1. Spatial shift and fusion}}
Let $p_{sum}=0$, $c=0$, $o = \mathrm{zeros}(x.\mathrm{shape})$\;
\For{offset $p$ in $D$}{
    calculate $d_p$ (Manhattan distance)\;
    calculate $w_p = 1/d_p$\;
    $p_{sum} \mathrel{+}= w_p$\;
}
calculate $k = C / p_{sum}$\;
\For{offset $p$ in $D$}{
    fill $o$ with shifted $x$: $o[c:c+k w_p] \gets x_p[c:c+k w_p]$\;
    $c \gets c+k w_p$\;
}
obtain shifted spatial $o_{spatial}$\;
\textbf{Fuse:} $s_{out} = \omega_{spatial}\cdot o_{spatial} + (1-\omega_{spatial})\cdot x$\;
\vspace{0.2em}
\tcp{\textbf{2. LL shift and fusion}}
Apply DWT to $x$ to obtain LL, LH, HL, HH\;
Repeat the above shift and fusion steps for LL to get $o_{LL}$\;
\textbf{Fuse:} $f_{out} = \omega_{LL}\cdot o_{LL} + (1-\omega_{LL})\cdot \mathrm{LL}$\;
Recombine $f_{out}$ with unshifted high-frequency components via inverse DWT to get $o_{freq}$\;
\textbf{Final output:} $o_{final} = s_{out} + o_{freq}$\;
\Return $\omega_{out}\cdot o_{final} + (1-\omega_{out})\cdot x$\;
\end{algorithm}

This design allows the network to better capture and propagate global contextual features critical for recovering broad tissue distributions and structural coherence. By applying shift operations exclusively to spatial and low-frequency branches, FSO-Shift strengthens the modeling of large-scale tissue contrast while avoiding the global information loss typical in conventional shift mechanisms. High-frequency components, which encode fine anatomical boundaries, are preserved throughout and directly fused back via inverse wavelet transform, maintaining structural integrity. A subsequent learnable fusion layer adaptively balances global and local information, enabling context-aware integration based on image content. 
\subsubsection{\textbf{Spatial Mix and Channel Mix}}
Each FS-RWKV block integrates spatial mix and channel mix modules to capture long-range spatial interactions and complex channel-wise dependencies. As shown in~\cref{fig:model}~(b), given image tokens $X \in \mathbb{R}^{T \times C}$ with $T=H \times W$, the spatial mix begins by applying the FSO-Shift operation, followed by three parallel linear projections to obtain:
\begin{equation}
\begin{aligned}
    R_s &= {FSO}\text{-}{Shift}_R(X)\, W_R, \\
    K_s &= {FSO}\text{-}{Shift}_K(X)\, W_K, \\
    V_s &= {FSO}\text{-}{Shift}_V(X)\, W_V,
\end{aligned}
\end{equation}
where $W_R$, $W_K$, and $W_V$ are learnable projection matrices. The shifted key and value tokens, $K_s$ and $V_s$, are passed to compute the global attention output, $wkv \in \mathbb{R}^{T \times C}$, via a linear complexity bidirectional attention mechanism, Bi-WKV~\cite{duan2025visionrwkv}. The attention calculation for the $t$-th token is calculated as follows:
\begin{equation}
\begin{aligned}
wkv_t &= {Bi}\text{-}{WKV}(K, V)_t \\
&= \frac{
    \sum_{\substack{i=0 \\ i \neq t}}^{T-1} e^{-\frac{|t - i| - 1}{T} \cdot w + k_i} v_i + e^{u + k_t} v_t
}{
    \sum_{\substack{i=0 \\ i \neq t}}^{T-1} e^{-\frac{|t - i| - 1}{T} \cdot w + k_i} + e^{u + k_t}
},
\end{aligned}
\end{equation}
where $w$, $k_i$, and $u$ are parameters controlling the decay and positional weights. This attention output is modulated by the sigmoid-activated receptance token, $\sigma(R_s)$, which gates the final spatial mix output $O_s$ as follows:
\begin{equation}
\begin{aligned}
O_s &= \bigl(\sigma(R_s) \odot wkv \bigr) W_O, \\
\quad wkv &= {Bi}\text{-}{WKV}(K_s, V_s),
\end{aligned}
\end{equation}
where $\odot$ denotes element-wise multiplication, and $W_O$ is a learnable projection matrix.

Subsequently, the tokens are passed into the channel-mix module for channel-wise feature fusion. Similar to the spatial mix, the channel mix obtains the receptance and key tokens as:
\begin{equation}
\begin{aligned}
R_c &= {FSO}\text{-}{Shift}_R(X) W_R, \\
K_c &= {FSO}\text{-}{Shift}_K(X) W_K,
\end{aligned}
\end{equation}
where $W_R$ and $W_K$ are learnable projection matrices. Within the channel-mix module, the value token $V_c$ is generated by applying a Squared ReLU activation on $K_c$ followed by a linear projection, expressed as:
\begin{equation}
V_c = {SquaredReLU}(K_c) W_V.
\end{equation}

The output $O_c$ is then computed by modulating $V_c$ with a gating mechanism controlled by the sigmoid-activated receptance token $\sigma(R_c)$, followed by a linear projection:
\begin{equation}
O_c = \bigl(\sigma(R_c) \odot V_c \bigr) W_O,
\end{equation}
where $W_V$, $W_O$ are learnable weights. This channel-wise fusion enables the network to model complex inter-channel dependencies, complementing the spatial mix to improve feature representation.

\subsection{Structural Fidelity Enhancement Block (SFEB)}

The Structural Fidelity Enhancement Block (SFEB) enhances tissue contrast and anatomical integrity by fusing complementary features extracted from both spatial and frequency domains. Given an input feature map $f_{in} \in \mathbb{R}^{H \times W \times C}$, SFEB first branches it into a spatial path and a frequency path.

In the spatial path, local structural features are extracted via a $3\times3$ convolution, followed by Layer Normalization (LN) and ReLU activation. These are then refined by the LSConv~\cite{wang2025lsnet}, which effectively models long-range spatial dependencies to maintain anatomical structure continuity and employs small-kernel aggregation to preserve fine anatomical boundaries:
\begin{equation}
    f_{spatial} = \text{LSConv}(\text{ReLU}(\text{LN}(\text{Conv}_{3\times3}(f_{in})))).
\end{equation}

In the frequency path, spectral and directional information is extracted by applying a discrete wavelet transform (DWT) on $f_{in}$, decomposing it into one low-frequency component $f_{LL} \in \mathbb{R}^{\frac{H}{2} \times \frac{W}{2} \times C}$ and three high-frequency components $f_{HF} = \{f_{LH}, f_{HL}, f_{HH}\}$:
\begin{equation}
    \{f_{LL}, f_{HF}\} = \text{DWT}(f_{in}).
\end{equation}

The low-frequency feature $f_{LL}$ is fed into multi-scale convolutional layers to capture semantic context at various resolutions:
\begin{equation}
\begin{aligned}
    f_{LL}^{3\times3} &= \text{Conv}_{3\times3}(f_{LL}), \\
    f_{LL}^{5\times5} &= \text{Conv}_{5\times5}(f_{LL}), \\
    f_{LL}^{7\times7} &= \text{Conv}_{7\times7}(f_{LL}).
\end{aligned}
\end{equation}

These are concatenated and refined using LSConv:
\begin{equation}
    f_{LL}^{refined} = \text{LSConv}(\text{Concat}(f_{LL}^{3\times3}, f_{LL}^{5\times5}, f_{LL}^{7\times7})).
\end{equation}
We adopt LSConv for low-frequency features due to its large-kernel receptive field, which is well-suited for modeling global tissue contrast and coarse anatomical structures.

Meanwhile, the high-frequency features $f_{LH}$, $f_{HL}$, and $f_{HH}$ are concatenated:
\begin{equation}
    f_{H} = \text{Concat}(f_{LH}, f_{HL}, f_{HH}),
\end{equation}
and then processed by parallel convolutions with multiple kernel sizes:
\begin{equation}
\begin{aligned}
    f_{H}^{3\times3} &= \text{Conv}_{3\times3}(f_{H}), \\
    f_{H}^{5\times5} &= \text{Conv}_{5\times5}(f_{H}), \\
    f_{H}^{7\times7} &= \text{Conv}_{7\times7}(f_{H}).
\end{aligned}
\end{equation}

Instead of LSConv, the fused high-frequency features are refined by a sequence of depthwise separable convolutions (DSConv) and layer normalization, which better preserves edge information and structural boundaries:
\begin{equation}
    f_{H}^{refined} = \text{DSConv}(\text{ReLU}(\text{LN}(\text{Concat}(f_{H}^{3\times3}, f_{H}^{5\times5}, f_{H}^{7\times7}))))
\end{equation}

The enhanced frequency-domain representation is reconstructed using the inverse DWT:
\begin{equation}
    f_{freq} = \text{IDWT}(f_{LL}^{refined}, f_{H}^{refined}).
\end{equation}

To adaptively integrate both paths, we apply global average pooling (GAP) and a gating network to compute scalar weights $w_{freq}$ and $w_{spatial}$ such that $w_{freq} + w_{spatial} = 1$:
\begin{equation}
    [w_{freq}, w_{spatial}] = \text{WeightNet}(\text{GAP}(f_{freq}, f_{spatial})).
\end{equation}

Finally, a weighted summation fuses both branches into the output:
\begin{equation}
    f_{out} = w_{freq} \cdot f_{freq} + w_{spatial} \cdot f_{spatial}.
\end{equation}
\begin{figure}[t]
    \centering
    \includegraphics[width=1\linewidth]{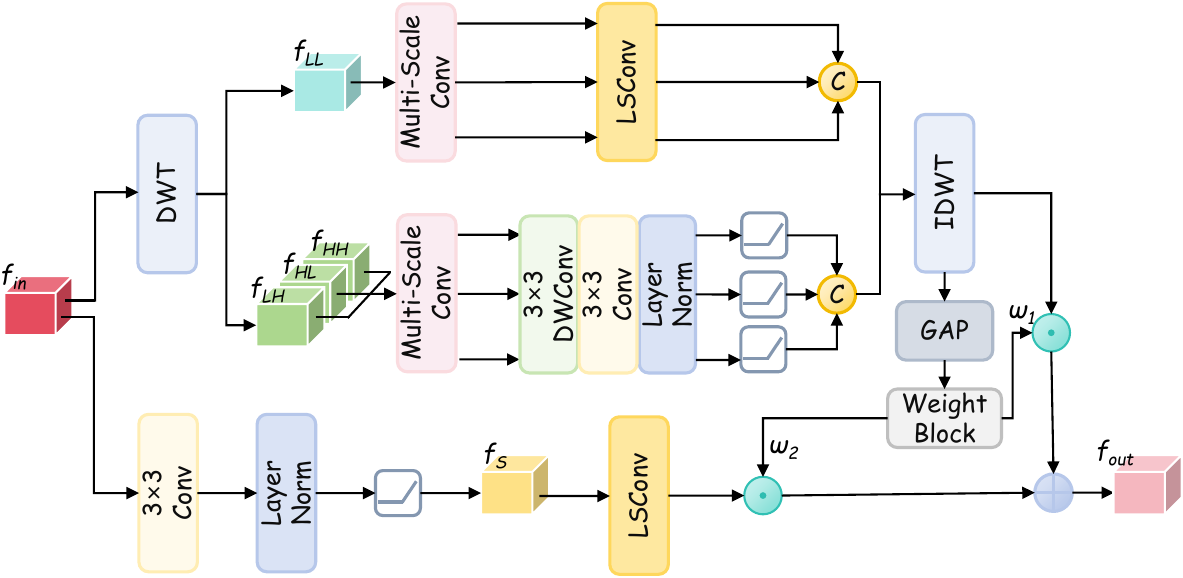}
    \caption{Overview of SFEB}
    \label{fig:vision_rwkv_architecture}
\end{figure}

\definecolor{cr2}{RGB}{175,255,197}
\definecolor{cr1}{RGB}{254,255,175}

\begin{table*}[htbp]
    \centering
    \caption{Quantitative comparison with state-of-the-art methods on the UNC and BNU datasets for 3T-to-7T T1w and T2w translation. The best results are \colorbox{cr1!70}{\textbf{bolded} and highlighted in yellow}, while the second-best results are \colorbox{cr2!70}{highlighted in green}.}
    \label{tab:unc_bnu_performance}
    \begin{minipage}[t]{0.49\linewidth}
    \centering
    \textbf{UNC Dataset}\\[2pt]
    \adjustbox{width=\linewidth}{
    \begin{tabular}{lccc|ccc}
    \toprule
    \multirow{2.5}{*}{\textbf{Method}} & \multicolumn{3}{c|}{\textbf{3T T1w $\rightarrow$ 7T T1w}} & \multicolumn{3}{c}{\textbf{3T T2w $\rightarrow$ 7T T2w}} \\
    \cmidrule{2-7}
    & PSNR (dB) $\uparrow$ & SSIM $\uparrow$ & RMSE $\downarrow$ & PSNR (dB) $\uparrow$ & SSIM $\uparrow$ & RMSE $\downarrow$ \\
    \midrule
    CycleGAN~\cite{zhu2017unpaired} & 19.9826 & 0.6789 & 0.1008 & \cellcolor{cr2!70}24.8817 & \cellcolor{cr2!70}0.7538 & \cellcolor{cr2!70}0.0586 \\
    pix2pix~\cite{isola2017image} & 19.9579 & 0.6847 & 0.1015 & 24.7376 & 0.7530 & 0.0591 \\
    DDCNN~\cite{zhang2019dual} & 20.7457 & 0.6412 & 0.0924 & 24.2051 & 0.7286 & 0.0640 \\
    WATNet~\cite{qu2020synthesized} & 20.1693 & 0.6753 & 0.0990 & 23.0667 & 0.6668 & 0.0719 \\
    RegGAN~\cite{kong2021breaking} & 19.4583 & 0.6423 &  0.1073 & 23.9740 & 0.6994 & 0.0654 \\
    ResViT~\cite{dalmaz2022resvit} & 20.1356 & 0.6696 & 0.0994 & 23.4681 & 0.7140 & 0.0687 \\
    PTNet~\cite{zhang2022ptnet3d} & \cellcolor{cr2!70}20.8183 & 0.7142 & \cellcolor{cr2!70}0.0916 & 24.6631 & 0.7495 &  0.0602 \\
    Resotre-RWKV~\cite{yang2025restore} & 20.3091 & \cellcolor{cr2!70}0.7158 & 0.0976 & 22.7059 & 0.7305 & 0.0781 \\
    \midrule
    \textbf{Ours} & \cellcolor{cr1!70}\textbf{21.0008} & \cellcolor{cr1!70}\textbf{0.7258} & \cellcolor{cr1!70}\textbf{0.0898} & \cellcolor{cr1!70}\textbf{25.3058} & \cellcolor{cr1!70}\textbf{0.7807} & \cellcolor{cr1!70}\textbf{0.0565} \\
    \bottomrule
    \end{tabular}}
    \end{minipage}%
    \hfill
    \begin{minipage}[t]{0.49\linewidth}
    \centering
    \textbf{BNU Dataset}\\[2pt]
    \adjustbox{width=\linewidth}{
    \begin{tabular}{lccc|ccc}
    \toprule
    \multirow{2.5}{*}{\textbf{Method}} & \multicolumn{3}{c|}{\textbf{3T T1w $\rightarrow$ 7T T1w}} & \multicolumn{3}{c}{\textbf{3T T2w $\rightarrow$ 7T T2w}} \\
    \cmidrule{2-7}
    & PSNR (dB) $\uparrow$ & SSIM $\uparrow$ & RMSE $\downarrow$ & PSNR (dB) $\uparrow$ & SSIM $\uparrow$ & RMSE $\downarrow$ \\
    \midrule
    CycleGAN~\cite{zhu2017unpaired} & 21.9260 & 0.7881 & 0.0807 & 26.3031 &  0.8053 & 0.0493 \\
    pix2pix~\cite{isola2017image} & 22.7388 & 0.8102 & 0.0739 & 26.4733 & 0.8182 & 0.0481 \\
    DDCNN~\cite{zhang2019dual} & 23.0212 & 0.7732 & 0.0717 & 26.8804 & 0.8229 & 0.0463 \\
    WATNet~\cite{qu2020synthesized} & 21.1316 &  0.7319 & 0.0886 & 26.1344 & 0.7625 & 0.0503 \\
    RegGAN~\cite{kong2021breaking} & 22.0967 & 0.7897 & 0.0793 & 25.6141 & 0.7703 & 0.0531 \\
    ResViT~\cite{dalmaz2022resvit} & 22.7787 & 0.8069 & 0.0733 & 26.5809 &  \cellcolor{cr2!70}0.8251 & 0.0477 \\
    PTNet~\cite{zhang2022ptnet3d} & 22.9820 & 0.7903 & 0.0717 & \cellcolor{cr2!70}27.3448 & 0.8242 & \cellcolor{cr2!70}0.0439 \\
    Resotre-RWKV~\cite{yang2025restore}& \cellcolor{cr2!70}23.3481 & \cellcolor{cr2!70}0.8366 & \cellcolor{cr1!70}\textbf{0.0688} & 24.4077 & 0.7676 & 0.0670 \\
    \midrule
    \textbf{Ours} & \cellcolor{cr1!70}\textbf{23.3571} & \cellcolor{cr1!70}\textbf{0.8388} & \cellcolor{cr2!70}0.0689 & \cellcolor{cr1!70}\textbf{27.4937} & \cellcolor{cr1!70}\textbf{0.8624} & \cellcolor{cr1!70}\textbf{0.0431} \\
    \bottomrule
    \end{tabular}}
    \end{minipage}
\end{table*}

\subsection{Loss Function}

To ensure both perceptual fidelity and anatomical accuracy in the synthesized 7T MR images, we employ a composite loss that combines pixel-wise, structural, and edge-aware supervision. The total loss is formulated as:
\begin{equation}
\mathcal{L}_{{total}} = \mathcal{L}_{{L1}} + \lambda_{{SSIM}} \cdot \mathcal{L}_{{SSIM}} + \lambda_{{edge}} \cdot \mathcal{L}_{{edge}}.
\end{equation}

The smooth L1 loss is computed as:
\begin{equation}
\begin{aligned}
\mathcal{L}_{{L1}} = \frac{1}{N} \sum_{i=1}^{N}
\begin{cases}
0.5 (d_i)^2, & \text{if } |d_i| < 1 \\
|d_i| - 0.5, & \text{otherwise}
\end{cases}, \\
\end{aligned}
\end{equation}
where $d_i = I_{{pred}}^{(i)} - I_{{gt}}^{(i)}$.
the SSIM loss is defined as:
\begin{equation}
\mathcal{L}_{{SSIM}} = 1 - {SSIM}(I_{{pred}}, I_{{gt}}),
\end{equation}
and the edge loss is given by:
\begin{equation}
\mathcal{L}_{{edge}} = \left\| \nabla I_{{pred}} - \nabla I_{{gt}} \right\|_1,
\end{equation}
with $\nabla$ denoting Sobel-based gradient magnitude. We empirically set $\lambda_{{SSIM}} = 0.4$ and $\lambda_{{edge}} = 0.3$ in all experiments. This loss formulation encourages the model to recover both global tissue contrast and fine anatomical detail while maintaining perceptual coherence and structural fidelity.

\section{Experiment}

\subsection{Paired 3T-7T MRI Dataset}
In our experiments, we utilized two publicly available datasets from~\cite{chen2023paired} and~\cite{chu2025paired}, which we refer to as the UNC Dataset and the BNU Dataset, respectively. 

\subsubsection{\textbf{Dataset Description}}
The UNC Dataset contains 10 participants aged 25--41 years. Brain MRI scans were obtained using 3T Siemens Magnetom Prisma and 7T Siemens Magnetom Terra scanners, both equipped with 32-channel head coils. The 3T acquisition protocol employed magnetization-prepared rapid gradient-echo (MPRAGE) for T1-weighted imaging (208 sagittal slices, repetition time (TR) = 2,400~ms, echo time (TE) = 2.2~ms, flip angle (FA) = $8^\circ$, acquisition matrix = $320 \times 320$, resolution = $0.8 \times 0.8 \times 0.8$~mm$^3$), while T2-weighted images were captured using sampling perfection with application-optimized contrasts using different flip angle evolution (SPACE) (208 sagittal slices, TR = 3,200~ms, TE = 563~ms, acquisition matrix = $320 \times 320$, resolution = $0.8 \times 0.8 \times 0.8$~mm$^3$). The 7T acquisition utilized magnetization-prepared 2 rapid acquisition gradient echoes (MP2RAGE) for T1-weighted imaging (256 sagittal slices, TR = 6,000~ms, TE = 1.91~ms, FA$_1$ = $4^\circ$, FA$_2$ = $4^\circ$, acquisition matrix = $304 \times 308$, resolution = $0.65 \times 0.65 \times 0.65$~mm$^3$), and turbo spin echo (TSE) for T2-weighted imaging (252 sagittal slices, TR = 3,000~ms, TE = 282~ms, FA$_1$ = $120^\circ$, FA$_2$ = $7^\circ$, acquisition matrix = $320 \times 208$, resolution = $0.65 \times 0.65 \times 0.65$~mm$^3$).

The BNU Dataset includes 20 participants aged 18--25 years. Paired 3T--7T MRI acquisitions were performed using a 3T Prisma scanner and a 7T MAGNETOM scanner. The 3T imaging protocol incorporated an isotropic 1~mm$^3$ sagittal 3D MPRAGE sequence for T1-weighted images (TR/inversion time/TE (TR/TI/TE) = 2,300/1,000/2.26~ms, flip angle = $8^\circ$, acquisition matrix size = $256 \times 224 \times 192$), complemented by TSE acquisition for T2-weighted images with a resolution of $0.9 \times 0.9 \times 1.9$~mm$^3$ (TR/TE = 11,050/94~ms, flip angle = $60^\circ$, bandwidth = 100~Hz/Px, acquisition matrix size = $256 \times 256 \times 90$). The 7T imaging employed a high-resolution isotropic 0.7~mm$^3$ sagittal 3D MPRAGE sequence for T1-weighted images (TR/TI/TE = 2,600/1,050/2.72~ms, flip angle = $5^\circ$, acquisition matrix size = $320 \times 320 \times 256$), alongside a 3D SPACE sequence for T2-weighted images with a resolution of $0.4 \times 0.4 \times 1.0$~mm$^3$ (TR/TE = 3,000/387~ms, flip angle = $120^\circ$, bandwidth = 279~Hz/Px, acquisition matrix size = $480 \times 512 \times 224$).

\subsubsection{\textbf{Dataset Preprocessing}}
For the UNC dataset, 3T T1w and 7T T2w images were linearly registered to the 7T T1w reference, while 3T T2w images were co-registered with the aligned 3T T1w to ensure spatial consistency. In the BNU dataset, both 3T and 7T T2w images were rigidly aligned to their respective T1w counterparts. The 7T T1w images were then linearly registered to the 3T T1w space, followed by alignment to a common unbiased template. All registration steps were performed using FSL FLIRT~\cite{jenkinson2002improved,jenkinson2012fsl}. After alignment, image intensities were normalized to $[0,1]$, and histogram matching with linear adjustment was applied to harmonize intensity distributions across modalities and datasets.
\begin{figure*}[ht]
    \centering
    \includegraphics[width=\linewidth]{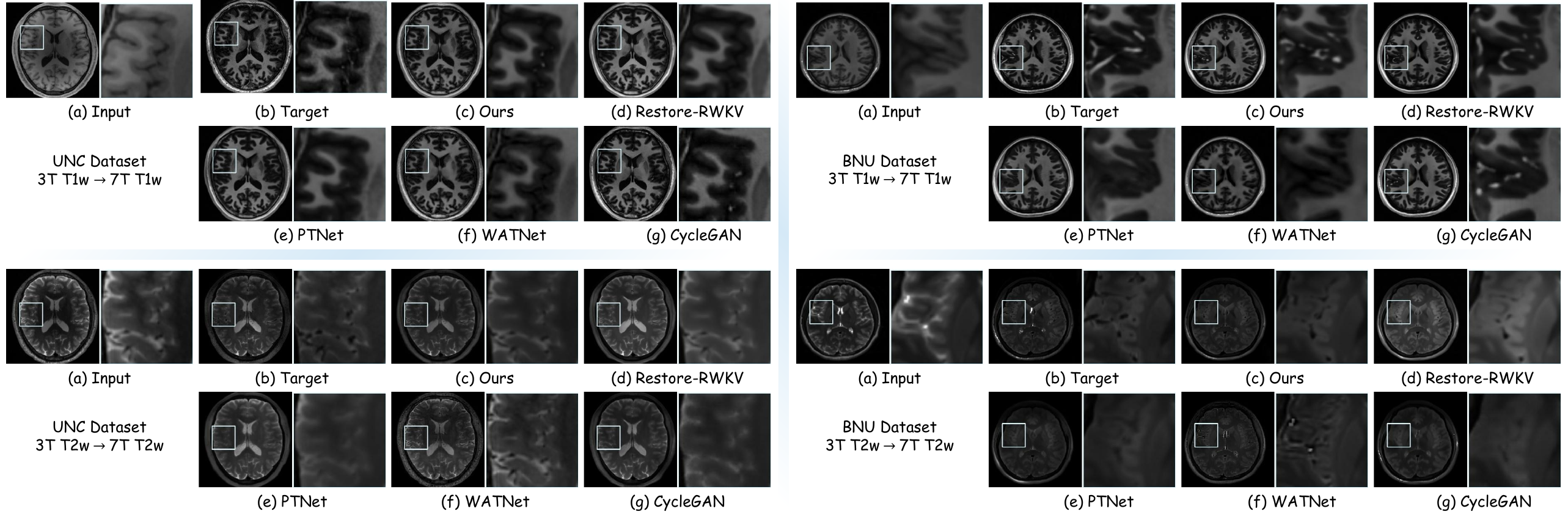}
    \caption{Qualitative comparison on UNC and BNU datasets for both T1w and T2w tasks. Our method better restores fine anatomical details and produces perceptually closer results to the 7T ground truth.}
    \label{fig:qualitative}
\end{figure*}
\subsection{Experimental Setup}
\subsubsection{\textbf{Implementation Details}}
All experiments were conducted using PyTorch on Ubuntu 22.04 with two NVIDIA RTX 3090 GPUs. We used a batch size of 1 and the AdamW~\cite{loshchilov2018decoupled} optimizer with an initial learning rate of $2 \times 10^{-4}$, $\beta_1 = 0.9$, $\beta_2 = 0.999$, and cosine annealing scheduling~\cite{loshchilov2017sgdr}. For preprocessing, we extracted 101 central axial slices per subject and resized all images to $256 \times 256$. The BNU dataset included 16 training and 4 test subjects (1,616/404 slices), while UNC had 7 training and 3 test subjects (707/303 slices). Data augmentation included random cropping, flipping, rotation, and mixup. All results are reported at $256 \times 256$ resolution.

\subsubsection{\textbf{Evaluation Metrics}}
To evaluate the quality of 7T synthesis, we adopt three widely used metrics: Peak Signal-to-Noise Ratio (PSNR), Structural Similarity Index Measure (SSIM)~\cite{wang2004image}, and Root Mean Squared Error (RMSE). PSNR measures global fidelity, SSIM evaluates perceptual and structural similarity, and RMSE captures pixel-wise reconstruction error.
\subsection{Comparisons with State-of-the-arts}
\begin{table}[b]
    \centering
    \caption{Ablation study results on the UNC dataset for both T1w and T2w synthesis.}
    \label{tab:ablation-joint}
    \adjustbox{width=0.95\linewidth}{
    \begin{tabular}{cc|ccc|ccc}
        \toprule
        \multicolumn{2}{c|}{\textbf{Modules}} & 
        \multicolumn{3}{c|}{\textbf{3T T1w $\rightarrow$ 7T T1w}} &
        \multicolumn{3}{c}{\textbf{3T T2w $\rightarrow$ 7T T2w}} \\
        \cmidrule{3-8}
        SFEB & FSO-Shift & 
        PSNR (dB) $\uparrow$ & SSIM $\uparrow$ & RMSE $\downarrow$ &
        PSNR (dB) $\uparrow$ & SSIM $\uparrow$ & RMSE $\downarrow$ \\
        \midrule
        \color{ForestGreen}\Checkmark & \color{Red}\XSolidBrush  
        & 20.7343 & 0.7082 & 0.0926 & 25.1936 & 0.7760 & 0.0565 \\
        
        \color{Red}\XSolidBrush & \color{ForestGreen}\Checkmark  
        & 20.7653 & 0.7159 & 0.0922 & 25.2534 & 0.7764 & \textbf{0.0557} \\
        
        \color{ForestGreen}\Checkmark & \color{ForestGreen}\Checkmark
        & \textbf{21.0008} & \textbf{0.7258} & \textbf{0.0898} & \textbf{25.3058} & \textbf{0.7807} & 0.0565 \\
        \bottomrule
    \end{tabular}}
\end{table}

To comprehensively evaluate the effectiveness of our method, we compare it against a series of representative baselines, including CNN-based (DDCNN~\cite{zhang2019dual}, WATNet~\cite{qu2020synthesized}), Transformer-based (ResViT~\cite{dalmaz2022resvit}, PTNet~\cite{zhang2022ptnet3d}), GAN-based (CycleGAN~\cite{zhu2017unpaired}, pix2pix~\cite{isola2017image}, RegGAN~\cite{kong2021breaking}), and RWKV-based (Restore-RWKV~\cite{yang2025restore}) architectures.

Quantitative results on both the UNC and BNU datasets are reported in~\cref{tab:unc_bnu_performance}. Our model consistently achieves top or second-best performance across all metrics, outperforming prior methods in PSNR, SSIM, and RMSE for both T1w and T2w modalities. This demonstrates the robustness and generalizability of our method across domains and imaging protocols. In particular, the improvements in SSIM and RMSE reflect the model’s ability to preserve anatomical structures and suppress artifacts, confirming the benefit of our proposed FSO-Shift and SFEB modules.

Qualitative comparisons are shown in~\cref{fig:qualitative}, where we visualize representative results from each category of competing methods, including CycleGAN (GAN-based), WATNet (CNN-based), PTNet (Transformer-based), and Restore-RWKV (RWKV-based). Across both datasets and modalities, our method produces images that are visually closest to the target in terms of anatomical detail and intensity distribution. Notably, our results exhibit clearer cortical boundaries and sharper tissue textures, while reducing common artifacts such as over-smoothing (in PTNet), structural deformation (in CycleGAN), and detail loss (in WATNet). These visual improvements corroborate the quantitative gains and highlight the benefits of our FSO-Shift and SFEB designs in preserving fine-grained anatomy.

\subsection{Ablation Studies}

To assess the contribution of each proposed component, we perform ablation experiments on the UNC dataset for both T1w and T2w synthesis. As reported in Table~\cref{tab:ablation-joint}, removing either SFEB or FSO-Shift results in a consistent drop in PSNR, SSIM, and RMSE, highlighting their complementary benefits. FSO-Shift improves global contrast and contextual representation by leveraging low-frequency-guided token interaction, while SFEB enhances anatomical fidelity through multi-domain enhancement. The full model achieves the best performance, demonstrating the synergy of global context modeling and structural detail preservation.

\section{Conclusion}
We proposed FS-RWKV, a novel RWKV-based framework for synthesizing ultra-high-field 7T MRI from 3T acquisitions. Built upon the efficient and globally-aware RWKV architecture, FS-RWKV integrates two core components: Frequency Spatial Omnidirectional-Shift (FSO-Shift) and Structural Fidelity Enhancement Block (SFEB). FSO-Shift applies directional token shifts to wavelet-decomposed low-frequency components, enabling the model to better preserve global tissue contrast while maintaining the integrity of high-frequency anatomical boundaries. SFEB further refines the output by adaptively enhancing multi-scale structural representations across both frequency and spatial domains. Through this design, FS-RWKV effectively reconstructs both broad intensity distributions and fine anatomical details. Extensive experiments on the UNC and BNU datasets demonstrate that FS-RWKV consistently outperforms state-of-the-art CNN-, Transformer-, GAN-, and RWKV-based baselines across T1w and T2w modalities. These results highlight the practical value of FS-RWKV in producing high-fidelity 7T-like images from 3T input, offering a scalable and generalizable solution that supports advanced neuroimaging and clinical analysis without requiring ultra-high-field acquisition.

\bibliography{ref}
\bibliographystyle{IEEEtran}
\end{document}